\documentclass[a4paper,11pt]{article}
\usepackage{pos}

\usepackage{graphicx,epsfig,amsmath}
\usepackage{subcaption}
\usepackage{multirow}
\usepackage{slashed}
\usepackage{cleveref}
\usepackage{enumitem}
\usepackage{calc}

\def\be{\begin{equation}}
\def\ee{\end{equation}}
\def\bea{\begin{eqnarray}}
\def\eea{\end{eqnarray}}

\newcommand{\mhh}{m_{hh}}

\title{Beyond dimension six in SM Effective Field Theory: \\a case study in Higgs pair production at NLO QCD}
\ShortTitle{Higgs pair production in SMEFT at NLO QCD: truncation uncertainties}

\author*[a]{Gudrun Heinrich}
\author[a]{Jannis Lang,}
\author[b]{Ludovic Scyboz}

\affiliation[a]{Institute for Theoretical Physics, Karlsruhe Institute of Technology (KIT),\\
76131 Karlsruhe, Germany}

\affiliation[b]{Rudolf Peierls Centre for Theoretical Physics, Parks Road, University of Oxford,\\
Oxford OX1 3PU, UK}

\emailAdd{gudrun.heinrich@kit.edu}
\emailAdd{jannis.lang@kit.edu}
\emailAdd{ludovic.scyboz@physics.ox.ac.uk}

\abstract{We present the NLO (two-loop) QCD corrections to Higgs boson pair production in gluon fusion within Standard Model Effective Theory (SMEFT), including also squared dimension-6 operators and double insertions of operators. The different options to truncate the EFT expansion are contrasted to a non-linear EFT approach (HEFT) and their effects are illustrated by several phenomenological examples.}

\FullConference{%
  Loops and Legs in Quantum Field Theory - LL2022,\\
  25-30 April, 2022\\
  Ettal, Germany
}

\begin{document}
\maketitle

\section{Introduction}

Higgs-boson pair production is commonly viewed as the golden channel to improve
our understanding of the Higgs potential.
The gluon fusion production channel is the process with the largest cross section 
that provides direct constraints on the Higgs trilinear self-coupling. The
effects of anomalous couplings are typically studied in the context of Effective
Field Theories (EFTs): In general, this procedure introduces a new theoretical
systematic uncertainty, associated with the truncation of the series in
the EFT expansion parameter. It is thus important to be able to reliably estimate of the  size
of these uncertainties.
Results for $gg \to hh$ have already been obtained in the non-linear Higgs Effective Field Theory
(HEFT)  at full NLO QCD~\cite{Buchalla:2018yce}, and later have been matched to parton
showers in a {\tt
Powheg-Box}~\cite{Nason:2004rx,Frixione:2007vw,Alioli:2010xd} implementation~\cite{Heinrich:2020ckp}.
Here, we present NLO QCD results for Higgs-boson pair production within the
Standard Model Effective Field Theory (SMEFT), based on the calculation explained in more detail in~\cite{Heinrich:2022idm}. Our
formalism allows the user to choose between different truncation options. The
NLO corrections are implemented in the public {\tt ggHH\_SMEFT} generator in the
{\tt Powheg-Box}.\footnote{Available at
https://powhegbox.mib.infn.it as User-Processes-V2/ggHH\_SMEFT.} 

\section{Effective field theory expansion schemes}
In this section, we define our conventions and present the EFT systematics we used. We focus on the linear Standard Model Effective field theory (SMEFT) expansion, and contrast it to the non-linear Higgs Effective Field Theory (HEFT, also called electroweak chiral Lagrangian).

The SMEFT~\cite{Buchmuller:1985jz,Grzadkowski:2010es,Brivio:2017vri} is an effective field theory with an expansion based on counting the canonical dimension of operators, where the Wilson coefficients of higher dimensional operators are suppressed by inverse powers of the scale $\Lambda$, a characteristic scale of the unknown new physics. Operators are composed of SM fields and SM symmetries are imposed, such that the Higgs field is contained in a doublet transforming linearly under  $SU(2)_L \times U(1)$.
In our study, we only consider the contributions of the leading dimension-6 operators and hence the Lagrangian is of  the form
\begin{equation}
\mathcal{L}_\text{SMEFT} = \mathcal{L}_\text{SM} + \sum_{i}\frac{C_i^{(6)}}{\Lambda^2}\mathcal{O}_i^{\rm{dim6}} +{\cal O}(\frac{1}{\Lambda^3})\; .
\label{eq:Ldim6}
\end{equation}
The number of independent operators at this level is already quite substantial, but only a restricted subset enters in gluon fusion Higgs pair production.

We work in the so-called Warsaw basis~\cite{Grzadkowski:2010es}, hence the relevant dimension-6 operators are given by
\begin{equation}
\begin{split}
\Delta\mathcal{L}_{\text{Warsaw}}&=\frac{C_{H,\Box}}{\Lambda^2} (\phi^{\dagger} \phi)\Box (\phi^{\dagger } \phi)+ \frac{C_{H D}}{\Lambda^2}(\phi^{\dagger} D_{\mu}\phi)^*(\phi^{\dagger}D^{\mu}\phi)+ \frac{C_H}{\Lambda^2} (\phi^{\dagger}\phi)^3 \\ &+\left( \frac{C_{uH}}{\Lambda^2} \phi^{\dagger}{\phi}\bar{q}_L\phi^c t_R + h.c.\right)+\frac{C_{H G}}{\Lambda^2} \phi^{\dagger} \phi G_{\mu\nu}^a G^{\mu\nu,a}\;.  \label{eq:warsaw}
\end{split}
\end{equation}
The chromo-magnetic operator is omitted since it comes with an additional loop-suppression factor  relative to the operators entering eq.~\eqref{eq:warsaw}~\cite{Buchalla:2022vjp,Buchalla:2018yce,Arzt:1994gp}.

In HEFT~\cite{Feruglio:1992wf,Burgess:1999ha,Grinstein:2007iv,Contino:2010mh,Alonso:2012px,Buchalla:2013rka} the chiral dimension $d_\chi$~\cite{Weinberg:1978kz} is used for the classification of operators, which is formally identical to a counting in loop orders with $d_\chi=2L+2$~\cite{Buchalla:2013eza,Krause:2016uhw}. The expansion parameter is given by $\frac{f^2}{\Lambda^2}\sim \frac{1}{16\pi^2}$, thus the Lagrangian can be expressed as
\begin{align}
  {\cal L}_{d_\chi}={\cal L}_{(d_\chi=2)}+\sum_{L=1}^\infty\sum_i\left(\frac{1}{16\pi^2}\right)^L c_i^{(L)} O^{(L)}_i\;.
  \label{eq:loop_expansion}
  \end{align}
The relevant terms for $gg\to hh$ up to $d_\chi=4$ are 
\begin{align}
\Delta{\cal L}_{\text{HEFT}}=
-m_t\left(c_t\frac{h}{v}+c_{tt}\frac{h^2}{v^2}\right)\,\bar{t}\,t -
c_{hhh} \frac{m_h^2}{2v} h^3+\frac{\alpha_s}{8\pi} \left( c_{ggh} \frac{h}{v}+
c_{gghh}\frac{h^2}{v^2}  \right)\, G^a_{\mu \nu} G^{a,\mu \nu}\;.
\label{eq:ewchl}
\end{align}
The anomalous couplings are a priori unrelated, since the physical Higgs field enters as a singlet under SM symmetries.

After expansion of the Higgs doublet in eq.~\eqref{eq:warsaw} around its vacuum expectation value and application of the field redefinition
\begin{align}
h\to h+v^2\frac{C_{H,\textrm{kin}}}{\Lambda^2}\left(h+\frac{h^2}{v}+\frac{h^3}{3v^2}\right)\;,\label{eq:field_redefinition}
\end{align}
with $$C_{H,\textrm{kin}}:=C_{H,\Box}-\frac{1}{4}\,C_{HD}\;,$$
the Higgs kinetic term acquires its canonical form (up to ${\cal O}\left(\Lambda^{-4}\right)$ terms).
Comparing terms in the Lagrangian with eq.~\eqref{eq:ewchl}, we end up with the translation of coupling coefficients listed in Table \ref{tab:translation}, valid at $\mathcal{O}(\Lambda^{-2})$ at the level of the Lagrangian. However, this translation has to be considered with care, since in SMEFT the EFT expansion is based on the assumption that $C_i \frac{s}{\Lambda^2}$ is a small quantity, allowing only for small deviations from the SM, whereas in HEFT the anomalous couplings $c_i$ can be of order ${\cal O}(1)$.

\begin{table}[htb]
\begin{center}
\begin{tabular}{ |c |c| }
\hline
HEFT&Warsaw\\
\hline
$c_{hhh}$ & $1-2\frac{v^2}{\Lambda^2}\frac{v^2}{m_h^2}\,C_H+3\frac{v^2}{\Lambda^2}\,C_{H,\textrm{kin}}$ \\
\hline
$c_t$ &  $1+\frac{v^2}{\Lambda^2}\,C_{H,\textrm{kin}} - \frac{v^2}{\Lambda^2} \frac{v}{\sqrt{2} m_t}\,C_{uH}$\\
\hline
$ c_{tt} $ & $-\frac{v^2}{\Lambda^2} \frac{3 v}{2\sqrt{2} m_t}\,C_{uH} + \frac{v^2}{\Lambda^2}\,C_{H,\textrm{kin}}$\\
\hline
$c_{ggh}$ &  $\frac{v^2}{\Lambda^2} \frac{8\pi }{\alpha_s} \,C_{HG}$ \\
\hline
$c_{gghh}$ &  $\frac{v^2}{\Lambda^2}\frac{4\pi}{\alpha_s} \,C_{HG}$ \\
\hline
\end{tabular}
\end{center}
\caption{Translation at Lagrangian level between operators in HEFT and SMEFT in the Warsaw basis.\label{tab:translation}}
\end{table}

The SMEFT series is usually truncated at order $\mathcal{O}(1/\Lambda^2)$, and
contributions from squared dimension-6 operators, as well as dimension-8
operators, are typically neglected since they are formally suppressed (of order
$\mathcal{O}(1/\Lambda^4$)). In order to study parts of the neglected
contributions, in the following we allow all possible dimension-6 operator
insertions at amplitude level. The amplitude can be separated into a pure SM
term (no insertion), a single insertion (dim6), and a double operator-insertion
(dim6$^2$) term, as shown  in eq.~\eqref{eq:amplitude_expansion}.

\begin{align}
{\cal M}=
&\ 
\vcenter{\hbox{\includegraphics[page=1,scale=0.9]{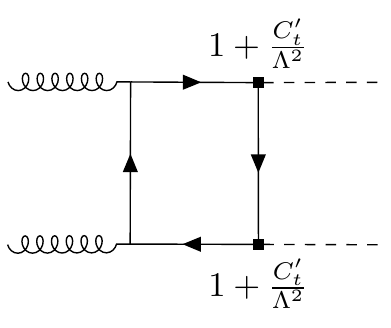}}}
+\vcenter{\hbox{\includegraphics[page=2,scale=0.9]{gghh_diagrams}}}
+\vcenter{\hbox{\includegraphics[page=3,scale=0.9]{gghh_diagrams}}}
\nonumber\\
&\ +\vcenter{\hbox{\includegraphics[page=4,scale=0.9]{gghh_diagrams}}}
+\vcenter{\hbox{\includegraphics[page=5,scale=0.9]{gghh_diagrams}}}
+\dots
\nonumber\\
=&\ 
{\cal M}_\text{SM} + {\cal M}_{\rm{dim6}} + {\cal M}_{\rm{dim6}^2}\;.\label{eq:amplitude_expansion}
\end{align}

When taking the square of the amplitude, $\sigma \propto |\mathcal{M}|^2$, we
define the following four truncation options, which differ in the way the above
terms are taken into account:

\begin{align}
\sigma \simeq \left\{\begin{aligned}
&\ \sigma_\text{SM} + \sigma_{\text{SM}\times \rm{dim6}} &\textrm{(a)}
\\
&\ \sigma_{\left(\text{SM}+\rm{dim6}\right)\times \left(\text{SM}+\rm{dim6}\right)}  &\textrm{(b)}
\\
&\ \sigma_{\left(\text{SM}+\rm{dim6}\right)\times \left(\text{SM}+\rm{dim6}\right)}  + \sigma_{\text{SM}\times \rm{dim6}^2} &\textrm{(c)}
\\
&\ \sigma_{\left(\text{SM}+\rm{dim6}+\rm{dim6}^2\right)\times \left(\text{SM}+\rm{dim6}+\rm{dim6}^2\right)} &\textrm{(d)}
\end{aligned}\right.\label{eq:truncation}
\end{align}

Option (a) corresponds to the first order of the expansion of $\sigma \propto
|\mathcal{M}|^2$ in $\Lambda^{-2}$. Option (b) is the first order of the
expansion of the amplitude $\mathcal{M}$ in $\Lambda^{-2}$. Option (c) includes
all terms stemming from single- and double-insertions of dimension-6 operators
at order $\mathcal{O}(\Lambda^{-4})$, i.e.~all contributions but those from
dimension-8 operators. Option (d) does not include any linearisation whatsoever,
thus corresponds to the case of HEFT upon translation of the parameters as given
in Table~\ref{tab:translation} (up to the running of the strong coupling
appearing in the effective gluon-Higgs couplings). In the following, we investigate
whether differences between these options can serve as a proxy for estimating
uncertainties related to the truncation of higher-dimensional operators.

\section{Cross-section results}
Our results were produced for a centre-of-mass energy of 
$\sqrt{s}=13$\,TeV, where we used the PDF4LHC15{\tt\_}nlo{\tt\_}30{\tt\_}pdfas~\cite{Butterworth:2015oua}
parton distribution functions interfaced to our code via
LHAPDF~\cite{Buckley:2014ana}, along with the corresponding value for
$\alpha_s$.  The masses of the Higgs boson and the top quark have been fixed,
as in the virtual amplitude, to $m_h=125$\,GeV, $m_t=173$\,GeV and their widths
have been set to zero.
Jets are clustered with the anti-$k_T$ algorithm~\cite{Cacciari:2008gp} as
implemented in the FastJet package~\cite{Cacciari:2005hq,
Cacciari:2011ma}, with jet radius $R=0.4$ and a minimum transverse momentum
$p_{T,\mathrm{min}}^{\rm{jet}}=20$\,GeV. We set the renormalisation and factorisation
scales to $\mu_R=\mu_F=m_{hh}/2$.

We show results for the SM and three benchmark points, derived
originally in Ref.~\cite{Capozi:2019xsi} based on a clustering of the EFT phase
space into seven characteristic $m_{hh}$-shapes. The original benchmark points were
refined to accommodate more recent experimental
constraints~\cite{CMS:2020gsy,ATLAS:2021vrm}, as well as the linear SMEFT
relation $c_{ggh} = 2c_{gghh}$. These benchmark points, marked with a star, are given in Table~\ref{tab:benchmarks},
with the corresponding values of the SMEFT coefficients as obtained by the
translation of Table~\ref{tab:translation} at $\Lambda = 1$ TeV.

\begin{table}[htb]
  \begin{center}
    \begin{footnotesize}
\begin{tabular}{ |c|c|c|c|c|c||c|c|c|c|c| }
\hline
\begin{tabular}{c}
benchmark \\
\end{tabular} & $c_{hhh}$ & $c_t$ & $ c_{tt} $ & $c_{ggh}$ & $c_{gghh}$ & $C_{H,\textrm{kin}}$ & $C_{H}$ & $C_{uH}$ & $C_{HG}$ & $\Lambda$\\
\hline
SM & $1$ & $1$ & $0$ & $0$ & $0$ & $0$ & $0$ & $0$ & $0$ & $1\;$TeV\\
\hline
$1^\ast$ & $5.105$ & $1.1$ & $0$ & $0$ & $0$ & $4.95$ & $-6.81$ & $3.28$ & $0$ & $1\;$TeV\\
\hline
$3^\ast$ & $2.21$ & $1.05$ & $ -\frac{1}{3} $ & $0.5$ & 0.25 & $13.5$ & $2.64$ & $12.6$ & $0.0387$ & $1\;$TeV\\
\hline
$6^\ast$ & $-0.684$ & $0.9$ & $ -\frac{1}{6} $ & $0.5$ & $0.25$ & $0.561$ & $3.80$ & $2.20$ & $0.0387$ & $1\;$TeV\\
\hline
\end{tabular}
\end{footnotesize}
\end{center}
\caption{Benchmark points used for the total cross sections and the distributions of the invariant mass of the Higgs-boson pair,
cf.~Table~\ref{tab:sigmatot} and Fig.~\ref{fig:bpdistributions}. The value of $C_{HG}$ is determined using $\alpha_s(m_Z)=0.118$.
\label{tab:benchmarks}}
\end{table}

Inclusive cross-sections are summarised in Table~\ref{tab:sigmatot} for
truncation option (b) with $\Lambda=1$ TeV and $\Lambda=2$ TeV, along with
results from truncation option (a) and HEFT. As evidenced for the case of
benchmark point 1, the purely linear truncation option (a) can lead to
unphysical cross sections, which serves to conclude that this benchmark point is
not a valid SMEFT point at $\Lambda=1$ TeV.

\begin{table}[htb]
\begin{center}
\begin{tabular}{| c | c | c |c|c|c|}
\hline
  benchmark  &$\sigma_{\rm{NLO}}$[fb]  &K-factor & ratio to  SM & $\sigma_{\rm{NLO}}$[fb] & $\sigma_{\rm{NLO}}$[fb] \\
point  & option (b) & option (b) & option (b) & option (a) & HEFT\\
  \hline
  \hline
SM & 27.94$^{+13.7\%}_{-12.8\%}$  & 1.67 & 1 & - &-\\
  \hline
  \hline
\multicolumn{6}{|c|}{$\Lambda=1$\,TeV}\\
  \hline
  \hline
$1^\ast$ & 74.29$^{+19.8\%}_{-15.6\%}$  & 2.13 & 2.66 & -61.17 & 94.32\\
\hline
$3^\ast$ & 69.20$^{+11.7\%}_{-10.3\%}$ & 1.82   & 2.47 & 29.64 & 72.43\\
\hline 
$6^\ast$ & 72.51$^{+20.6\%}_{-16.4\%}$& 1.90 & 2.60 & 52.89 & 91.40\\
  \hline
  \hline
\multicolumn{6}{|c|}{$\Lambda=2$\,TeV}\\
  \hline
  \hline
$1^\ast$& 14.03$^{+12.0\%}_{-11.9\%}$  &1.56  & 0.502 & 5.58 & -\\
\hline
$3^\ast$& 30.81$^{+16.0\%}_{-14.4\%}$ & 1.71 & 1.10 & 28.35  & -\\
\hline 
$6^\ast$& 35.39$^{+17.5\%}_{-15.2\%}$ & 1.76 & 1.27 & 34.18 & -\\
  \hline
\end{tabular}
\end{center}
\caption{Total cross sections for Higgs-boson pair production at full NLO QCD
for three benchmark points and truncation option (b).  The total cross sections
for truncation option (a) (linearised dim-6) are also given, in order to highlight the difference, as well as the values for HEFT. The fact that
truncation option (a) leads to a negative cross section for benchmark 1 clearly indicates that this is not a valid parameter point in SMEFT for $\Lambda=1$\,TeV.  The uncertainties are scale uncertainties based on
3-point scale variations.
\label{tab:sigmatot}}
\end{table}

\begin{figure}[h]
\begin{center}
\includegraphics[width=.33\textwidth,page=1]{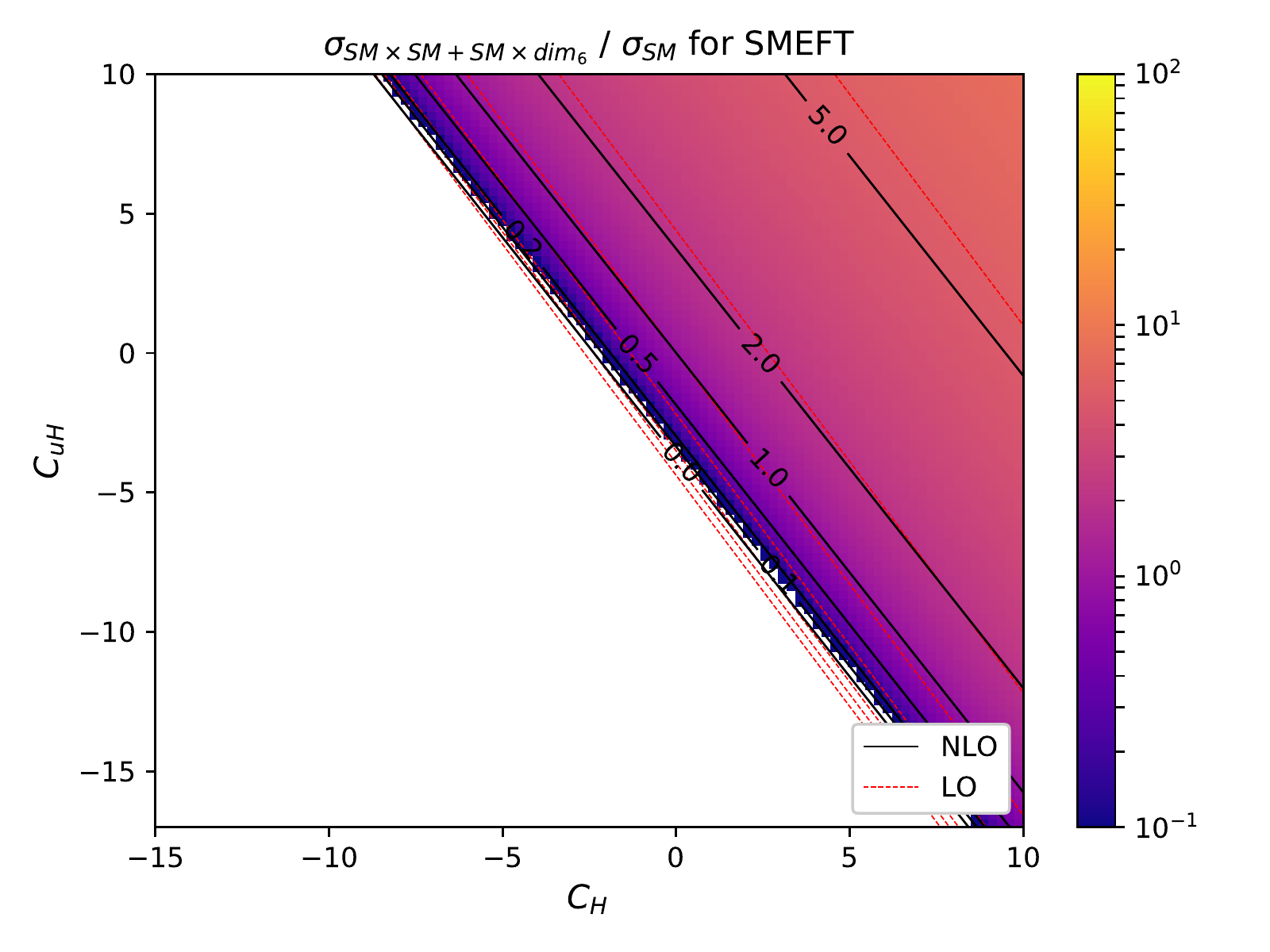}%
\includegraphics[width=.33\textwidth,page=2]{plot_CH_CuH.pdf}%
\includegraphics[width=.33\textwidth,page=4]{plot_CH_CuH.pdf}%
    \caption{\label{fig:heatmaps} Heat maps showing
    the dependence of the cross section on the couplings $C_H$, $C_{uH}$ (left)
    and $C_H$, $C_{H,\textrm{kin}}$ (right) with $\Lambda=1$\,TeV for different truncation
    options. Top: option (a) (linear dim-6), middle: option (b) (quadratic dim-6),
    bottom: option (d) (no linearisation in $1/\Lambda^2$).
    The white areas denote regions in parameter space where the corresponding cross
    section would be negative.}
\end{center}
\end{figure}

The ratio of the cross section to the SM value, $\sigma / \sigma_{\rm SM}$, is
shown as a function of the couplings $C_H$, $C_{uH}$ in Fig.~\ref{fig:heatmaps}
for the linear (left), quadratic (middle) and HEFT-like (right) truncation
options. A large part of the parameter space is characterised by negative
cross-sections for truncation option (a) (blank patch in the left plot). The
iso-contours of the cross-section become distorted by higher-order monomials in
the couplings, when going from the linearised case to the quadratic,
respectively the HEFT-like case.

%
%
%

\begin{figure}[h]
\includegraphics[width=17pc,page=1]{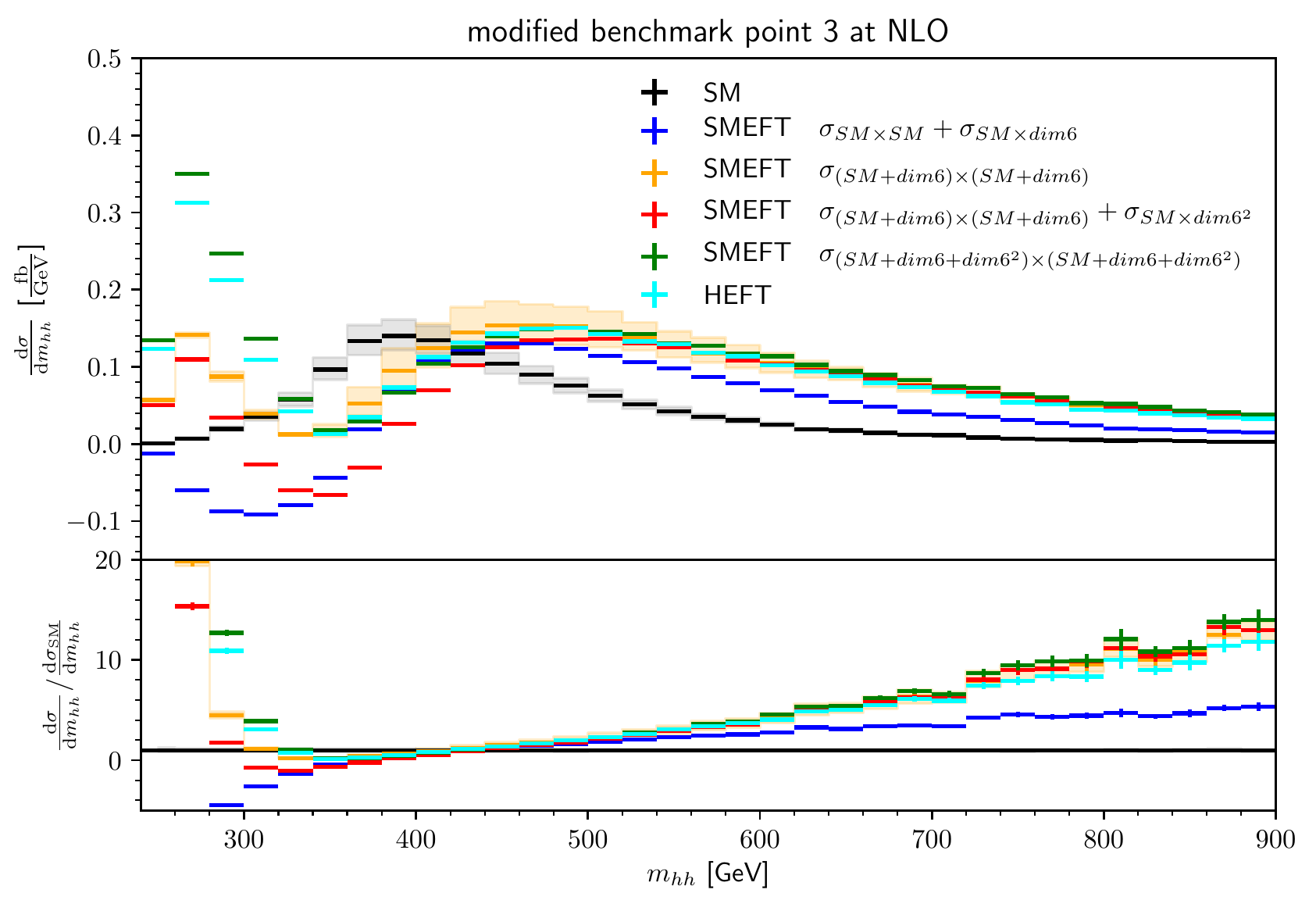}\hspace{2pc}%
\includegraphics[width=17pc,page=1]{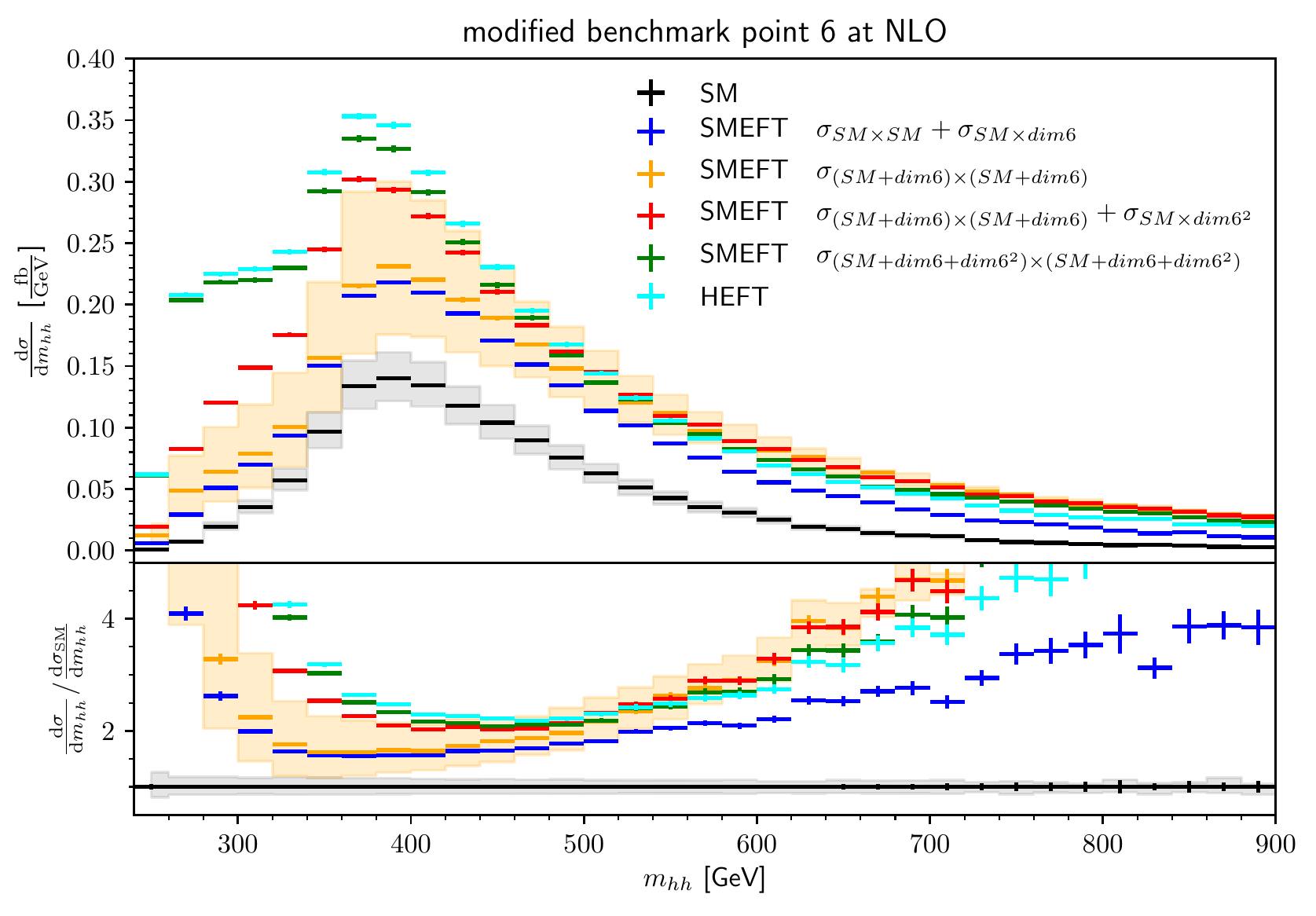}\hspace{2pc}%
\\
\includegraphics[width=17pc,page=1]{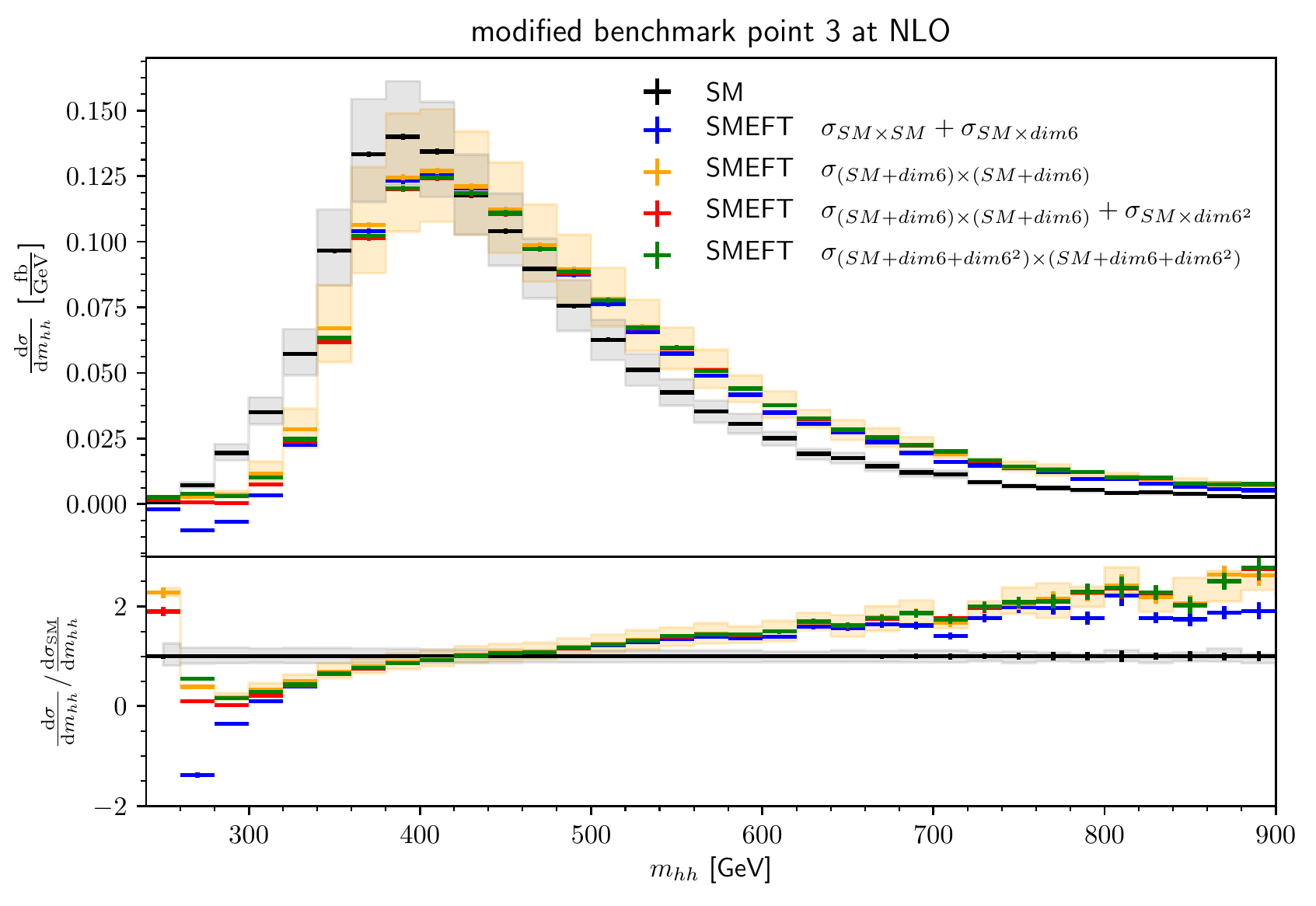}\hspace{2pc}%
\includegraphics[width=17pc,page=1]{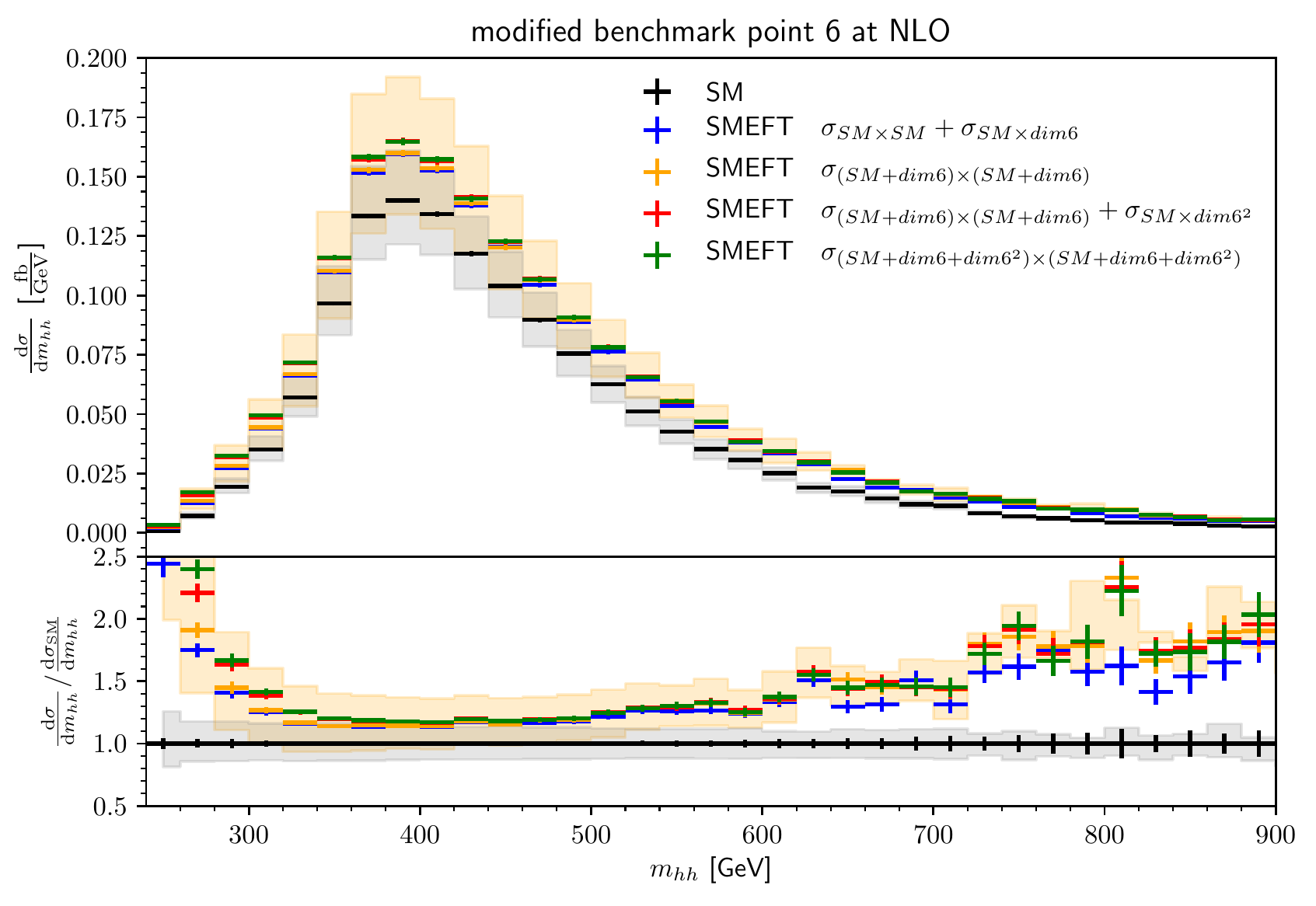}\hspace{2pc}%
\\
\includegraphics[width=17pc,page=1]{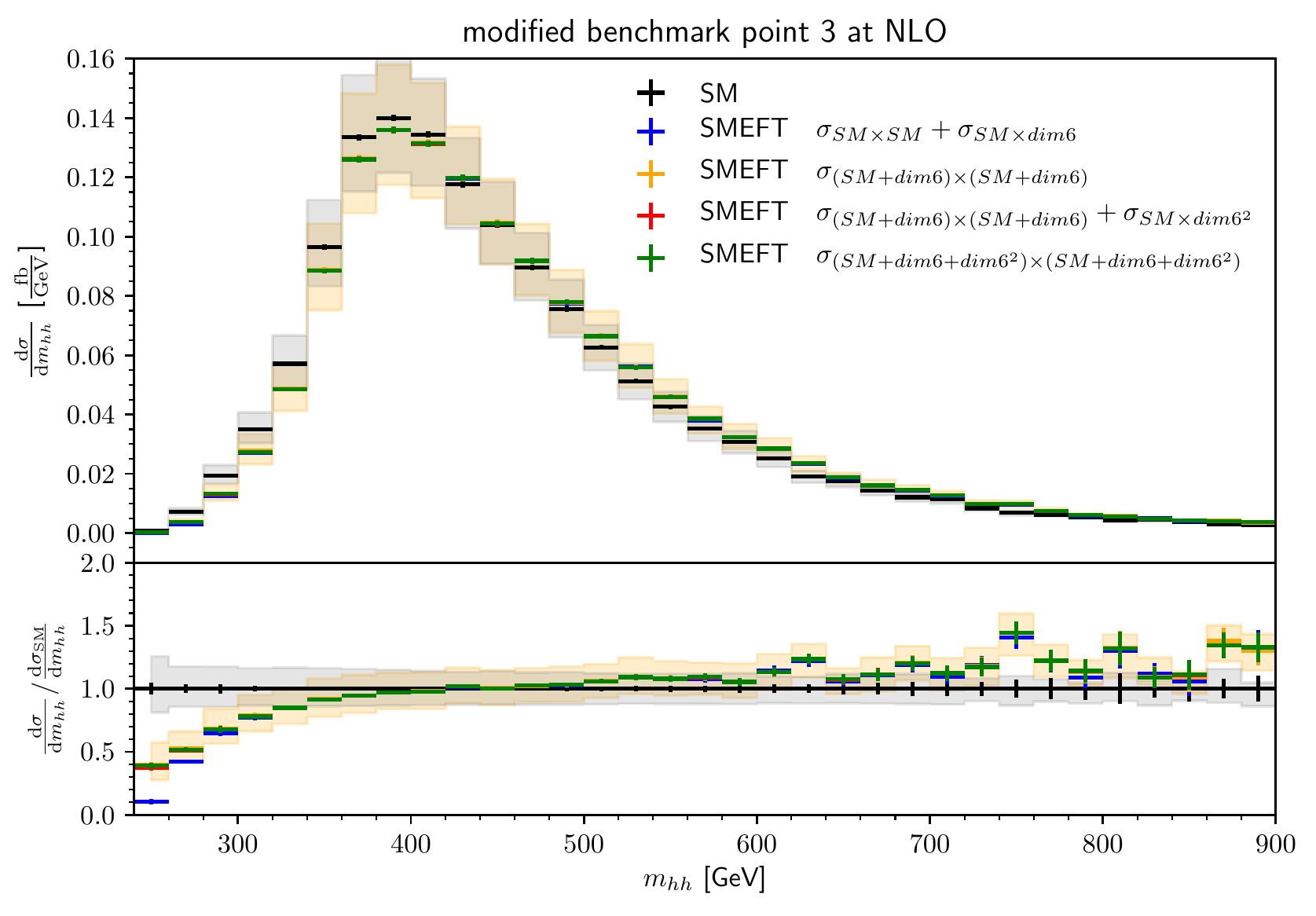}\hspace{2pc}%
\includegraphics[width=17pc,page=1]{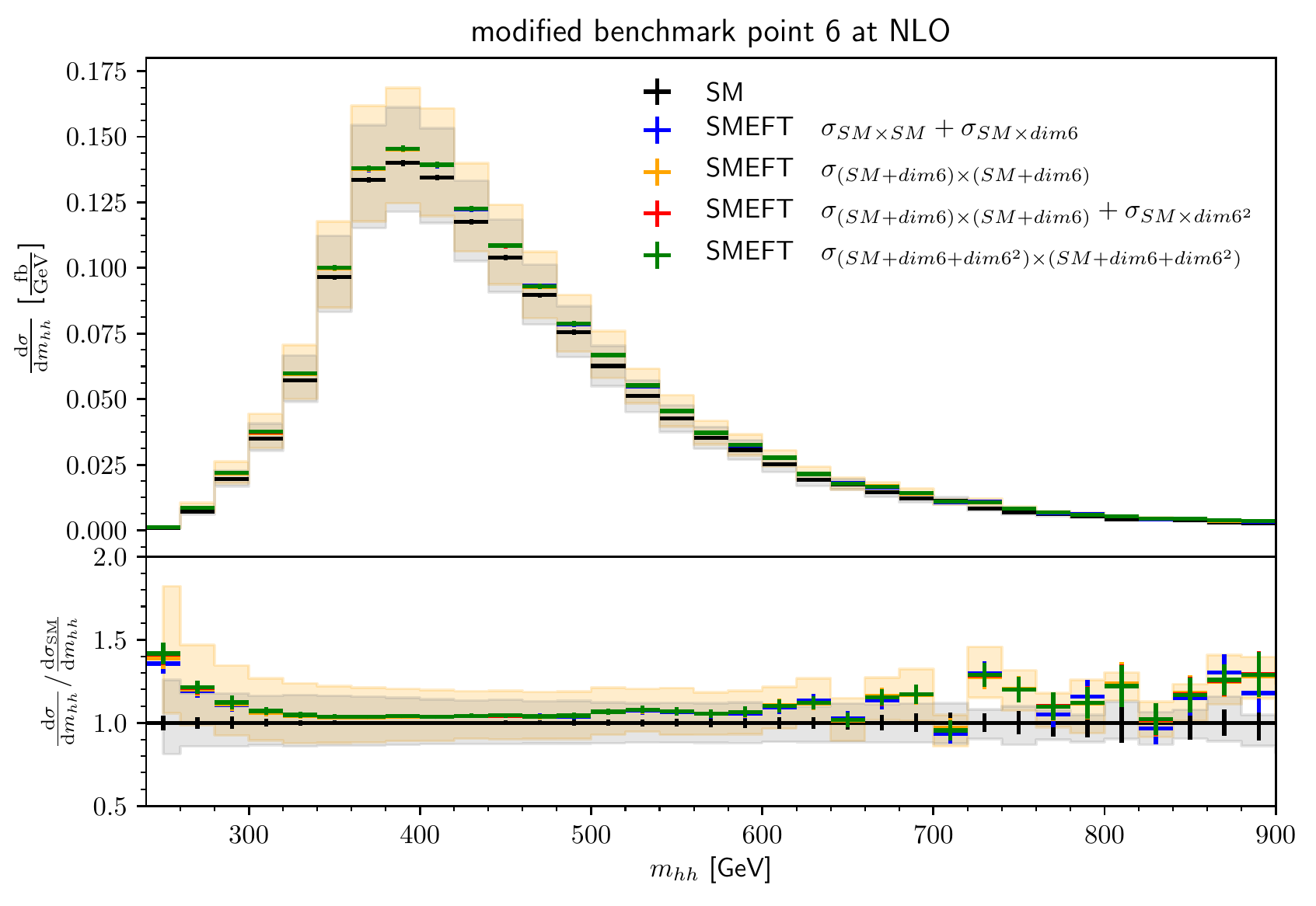}\hspace{2pc}%
   \caption{\label{fig:bpdistributions} Distribution of the invariant mass $m_{hh}$ of the
   Higgs-boson pair for two benchmark points of Table~\ref{tab:benchmarks}, with $\Lambda=1$\,TeV (top),
   $\Lambda=2$\,TeV (middle), and $\Lambda=4$\,TeV (bottom).  Left: benchmark 3$^\ast$, right: benchmark 6$^\ast$.}
\end{figure}

Finally, in Fig.~\ref{fig:bpdistributions} we present differential results for
the Higgs-pair invariant mass $m_{hh}$, for the benchmark
points 3 (left column) and 6 (right column) given in Table~\ref{tab:benchmarks},
at $\Lambda=1$\,TeV (top row), $\Lambda=2$\,TeV (middle row) and $\Lambda=4$\,TeV
(bottom row). As noted previously, truncation option (a) (dark blue) can lead to
unphysical cross-sections in part of the phase-space (here for benchmark point 3 at
$\Lambda\leq 2$ TeV). We show a 3-point scale variation around the central scale
$\mu_R = \mu_F = c \cdot m_{hh}/2$, with $c \in \lbrace \frac{1}{2}, 1,2
\rbrace$, for the SM curve (black) and truncation option (b) (orange). We also
include the HEFT curve (cyan) in the figures for $\Lambda=1$\,TeV for comparison.
Truncation option (d) (dark green) is formally equivalent to HEFT, and the
difference between both curves stems purely from the running of $\alpha_s$ in
front of the $C_{HG}$ coefficient. For both benchmarks, at $\Lambda=1$ TeV,
truncation options (b)--(d) retain -- if only marginally -- the characteristic
$m_{hh}$-shape identified in Ref.~\cite{Capozi:2019xsi} (double peak separated
by a dip in benchmark 3, and a shoulder left of the peak in benchmark 6).
Obviously, as the value of the
heavy scale $\Lambda$ is increased, the effect of the SMEFT operators becomes
numerically suppressed, and the differential distributions for all truncation options converge to the SM
curve.

\section{Conclusions}
We have presented the NLO QCD corrections to Higgs-boson pair production, with effects of BSM physics parametrised within the SMEFT framework, including  operators up to dimension-6. The calculation is implemented in the {\tt Powheg-Box-V2} in a flexible way, which allows different truncation options regarding multiple insertions of dimension-6 operators as well as the possibility to switch to the non-linear HEFT parametrisation for comparison.
The results show  that a naive translation from valid HEFT anomalous coupling parameter choices can lead to SMEFT parameters which are outside the validity range of the SMEFT expansion, as cross sections for SMEFT truncated at  linear dimension-6 level can turn negative.
Moreover, characteristic shapes of $\mhh$ distributions for HEFT benchmark points can only partly be recovered by some SMEFT truncation options after translation, since the interference patterns between the various contributions to the amplitude are different.
Points in the coupling parameter space that are clearly within the validity range of SMEFT tend to lead to only small distortions of the SM $\mhh$ distribution, such that they are within the scale uncertainty band of the SM case, and therefore emphasize the need for precise SM predictions.

\bibliographystyle{iopart-num}

\bibliography{HH_SMEFT_LL_main}


\end{document}